\documentclass{llncs}

\usepackage{lineno,hyperref}
\usepackage{listings}
\usepackage{xcolor}
\colorlet{punct}{red!60!black}
\definecolor{background}{HTML}{EEEEEE}
\definecolor{delim}{RGB}{20,105,176}
\colorlet{numb}{magenta!60!black}
\lstdefinelanguage{json}{
    basicstyle=\normalfont\ttfamily,
    numbers=left,
    numberstyle=\scriptsize,
    stepnumber=1,
    numbersep=8pt,
    showstringspaces=false,
    breaklines=true,
    frame=lines,
    backgroundcolor=\color{background},
}

\usepackage{booktabs}
\usepackage{graphicx}
\usepackage{amsmath}
\usepackage{amssymb} 
\usepackage{color}
\usepackage{mathtools}
\usepackage{proof-dashed}
\usepackage{mathpartir}
\usepackage{extarrows}
\usepackage[draft,nomargin,inline]{fixme} 
\usepackage{xcolor} 
\usepackage{tikz}
\usetikzlibrary{arrows,shapes,automata,spy,positioning,
  calc,shadows,fit,arrows.meta,patterns,backgrounds,decorations.pathreplacing}
\usepackage{tikz-qtree}
\usepackage{tabularx}
\usepackage{cmll} 
\usepackage[caption = false]{subfig}
\usepackage[export]{adjustbox}

\title{Tool paper title}

\title{A Session Subtyping Tool (Extended Version)}
\author{Lorenzo Bacchiani\inst{1}
 \ \ \  Mario Bravetti\inst{2}
 \ \ \   Julien Lange\inst{3}
 \ \ \  Gianluigi Zavattaro\inst{2}
}
\institute{University of Bologna, Italy \and
Department of Computer Science and Engineering \&
 Focus Team, INRIA \\ 
University of Bologna,  Italy \and 
Royal Holloway, University of London, Egham, UK
}

\date{}

\newcommand{\dual}[1]{\overline{#1}}

\newcommand{\infr}[2]
        {\renewcommand{\arraystretch}{1.5}
        \begin{array}{c}
        #1\\
        \hline
        #2
        \end{array}}

\newcommand{\auxarrow}
        {\mathop{\longrightarrow}}

\newcommand{\arrow}[1]
        {\, \auxarrow\limits^{#1} \,}

\newcommand{\Tbranchindex}[4]{\&\{{#1}_{#3};{#2}_{#3}\}_{#3\in #4}}
\newcommand{\Tbranch}[2]{\Tbranchindex{#1}{#2}{i}{I}}

\newcommand{\Tbra}[1]{\&\{#1\}}
\newcommand{\Tselectindex}[4]{\oplus\{{#1}_{#3};{#2}_{#3}\}_{#3\in #4}}
\newcommand{\Tselect}[2]{\Tselectindex{#1}{#2}{i}{I}}

\newcommand{\Tsel}[1]{\oplus\{ #1 \}}

\newcommand{\Trec}[1]{\mu \mathbf{#1}}
\newcommand{\Tvar}[1]{\mathbf{#1}}
\newcommand{\Tend}{\mathbf{end}}
\newcommand{\exTM}{\msg{tm}}
\newcommand{\exTC}{\msg{tc}}
\newcommand{\exDONE}{\msg{done}}
\newcommand{\exOVER}{\msg{over}}

\DeclareMathAlphabet{\mathpzc}{OT1}{pzc}{m}{it}

\newcommand{\calt}{{\cal T}}
\newcommand{\call}{{\cal L}}
\newcommand{\cala}{{\cal A}}

\newcommand{\cali}{{\cal I}}

\newcommand{\qst}{\, \mid \,}
\newcommand{\dst}{.\ }

\newcommand{\word}{\omega}

\newcommand{\act}{\ell}

\newcommand{\msg}[1]{\mathit{#1}}
\newcommand{\snd}[1]{!\msg{#1}}
\newcommand{\rcv}[1]{?\msg{#1}}
\newcommand{\trans}[1]{\xrightarrow{#1}}

\newcommand{\simtreetrans}[1]{\xhookrightarrow{#1}}

\newcommand{\States}{Q}

\newcommand{\state}{q}

\newcommand{\Trans}{\rightarrow} 

\newcommand{\sndL}[1]{! #1}
\newcommand{\rcvL}[1]{? #1}

\DeclareMathOperator{\nincsymbol}{\preccurlyeq}

\newcommand{\loopio}[2]{\mathsf{cycle}(#1,#2)}

\newcommand{\simtreepair}[2]{#1 \nincsymbol #2}
\newcommand{\ctx}[1]{\mathcal{#1}}

\newcommand{\echoice}[2]{\langle #1 \rangle_{#2}}

\newcommand{\inTrans}[1]{\mathsf{in}(#1)}
\newcommand{\outTrans}[1]{\mathsf{out}(#1)}
\newcommand{\intree}[1]{\mathsf{inTree}(#1)}

\newcommand{\exHQ}{\msg{nd}}
\newcommand{\exLQ}{\msg{pr}}
\newcommand{\exOK}{\msg{ok}}
\newcommand{\exKO}{\msg{ko}} 
\tikzset{
  every state/.style={minimum size=1pt,inner sep=1.5pt, initial text={}},
  mycfsm/.style={
    font=\scriptsize,
    initial where=left,
    initial distance=0.25cm,
    ->,>=stealth,auto, node distance=0.8cm and 0.8cm,
    scale=1, every node/.style={transform shape},
    baseline=(current  bounding  box.center)
  },
  ogate/.style = {
    diamond, draw, fill=white,
    minimum size=4mm,
    inner sep=0pt,
    postaction={path picture={%
        \draw[black]
        ([yshift=\gatedistancein]path picture bounding box.south) -- ([yshift=-\gatedistancein]path picture bounding box.north)
        ([xshift=-\gatedistancein]path picture bounding box.east) -- ([xshift=\gatedistancein]path picture bounding box.west)
        ;}}, drop shadow},
  agate/.style={draw,rectangle,
    minimum size=3mm,
    inner sep=0pt,
    fill=white,
    postaction={path picture={%
        \draw[black]
        ([yshift=\gatedistanceinand]path picture bounding box.south) --
        ([yshift=-\gatedistanceinand]path picture bounding box.north) ;}}, drop shadow},
  source/.style={draw,circle,fill=white,
    minimum size=3mm,
    inner sep=0pt, drop shadow},
  sink/.style={draw,circle,double,fill=white,
    minimum size=3mm,
    inner sep=0pt, drop shadow},
  intera/.style = {rectangle, draw=black, align=center, fill=white, rounded corners=0.1cm,
    minimum height=12,
    inner sep=2pt, drop shadow},
  line/.style = {draw,->, rounded corners=0.07cm,>=latex},
  venn/.style={preaction={fill, #1},opacity=0.6},
  cnode/.style={rectangle,draw=black,inner sep=2pt},
  ancestor/.style={densely dashed,->},
  hookedsilentedge/.style={>=latex,right hook->},
  lhookedsilentedge/.style={>=latex,left hook->},
  silentedge/.style={>=latex,->},
  nlabel/.style={fill=white,inner sep=0pt,font=\footnotesize},
  notexplo/.style={fill=gray!10},
  echnode/.style={rectangle,draw=black,inner sep=2pt},
  schnode/.style={diamond,draw=black,inner sep=0pt},
}

\DeclareMathOperator{\mysum}{\with}
\newcommand{\inchoicetop}{\oplus} 
\newcommand{\outchoicetop}{\mysum}

\newcommand{\inference}[3]{\infer[\ifthenelse{\equal{#1}{}}{}{\inferrule{#1}}]{#3}{#2}}
\newcommand{\coinference}[3]{\infer=[\ifthenelse{\equal{#1}{}}{}{\inferrule{#1}}]{#3}{#2}}

\begin{document}

\maketitle

\begin{abstract}
Session types are becoming popular and have been integrated in 
several mainstream programming languages. Nevertheless, 
while many programming languages consider asynchronous \textsc{fifo}
channel communication, the notion of subtyping used in session type implementations is 
the one defined by Gay and Hole for synchronous communication.
This might be because there are several notions of asynchronous session 
subtyping, these notions are usually undecidable, and only 
recently sound (but not complete) algorithmic characterizations
for these subtypings have been proposed. 
But the fact that the definition of asynchronous session subtyping and 
the theory behind related algorithms are not easily accessible to 
non-experts may also prevent further integration.  
The aim of this paper, and of the tool presented therein,
is to make the growing body of knowledge about asynchronous session subtyping more accessible,
thus promoting its integration in practical applications of session types.


%

\end{abstract}

\section{Introduction}

%
%
%
%

%

In recent years, session types have 
been integrated into several mainstream programming languages
(see, e.g.,~\cite{HuY16,Padovani17,SY2016,LindleyM16,OrchardY16,AnconaEtAl16,NHYA2018})
where they specify the pattern of interactions that each endpoint must
follow, i.e., a communication protocol.
All of these practical applications show 
a good level of maturity of the session type theory, 
but there are still some limitations. In particular, the notion of
subtyping considered in such tools usually assumes 
synchronous communication channels, 
while, in many cases, communication
takes place over asynchronous point-to-point \textsc{fifo} channels
(where outputs are non-blocking). In this setting, the emitted messages are stored 
inside channels, and there may be an arbitrary delay between 
an output (on an endpoint) and the corresponding input (on the opposite
endpoint).
The impact on session subtyping of these aspects related with 
asynchronous communication has been 
initially studied in \cite{MostrousESOP09,MostrousIC15,LMCSasync}, but the
notions of subtyping proposed therein were subsequently proved to be
undecidable \cite{BravettiCZ17,LangeY17}. Only recently, sound (but not complete)
algorithms for asynchronous session subtyping have been
proposed~\cite{BCZ18,BCLYZ21,BLZ21}.
However, the theory behind asynchronous session types (see \cite{BravettiZ21} for a gentle introduction) and 
related algorithms is rather intricate and this  
could limit their dissemination in the research community, 
as well as their adoption in practical applications.

The aim of this paper, and of the tool that we introduce, 
is to make the growing body of knowledge about asynchronous session subtyping more accessible.
More precisely, we present in an uniform and intuitive way various 
notions of (a)synchronous session subtyping that were presented in the literature
following different formalisms, e.g., types in \cite{BLZ21} or 
communicating finite-state machines in \cite{BCLYZ21}. Our tool 
integrates several algorithms for checking subtyping that can 
be invoked from an easy-to-use Python GUI. This interface
allows the user to input, using standard session type
syntax, two types:
the candidate subtype and supertype. The tool
automatically generates the 
graphical representation of these session types 
as communicating finite-state
machines~\cite{cfsm83}. It is also possible
to execute on them the desired subtyping algorithm(s).
The tool has been implemented in a modular way, and it 
is possible to easily include several subtyping algorithms, 
simply by customizing a JSON configuration file. In the current
version, we consider: two algorithms from~\cite{LangeY16} for synchronous 
session subtyping (based on Gay and Hole's~\cite{GH05}  and  Kozen et al.'s~\cite{KozenPS95} algorithms), a sound algorithm 
for checking (orphan message free) asynchronous
session subtyping \cite{BCLYZ21}, and a sound algorithm for
checking fair asynchronous session subtyping \cite{BLZ21}.
The implementations of these algorithms, besides returning 
a verdict about subtyping of the two types, 
also return a graphical representation of the so-called
\emph{subtyping simulation game}: i.e., 
the procedure to check that each relevant input/output action that can be performed
by the candidate subtype has a corresponding matching action in the candidate supertype.
This graphical representation is helpful to 
understand the reason behind the given verdict.
The original command line Haskell implementations of the algorithms in \cite{LangeY16,BCLYZ21,BLZ21} have been adapted and integrated by: $(i)$ uniformizing their graphical notation/colors (e.g.,  {\it inner/outer} states represented as rectangles, with the initial one being thicker, error ones being red, content of outer ones being blue, etc\dots),  $(ii)$ reimplementing the synchronous algorithm so to also generate the simulation graph, $(iii)$ completely rewriting, in the fair asynchronous algorithm, the controllability check (existance of a compliant peer, see Section~\ref{examples}) and $(iv)$ error detection with generation of red states for all algorithms, $(v)$ pre-transforming inputted types with a Python ANTLR4 parser that produces a common raw syntax.

%
\textit{Synopsis.}
Section \ref{SEC:subtyping} recalls basic notions about session 
subtyping using tool-simulated examples 
and Section \ref{SEC:tool} describes the functionalities of the tool. Finally, in Section \ref{SEC:conc} we conclude the paper. 

The tool sources/binaries are available at~\cite{toolrepo}.

%
%

\section{Session Subtyping} \label{SEC:subtyping}

\begin{figure}[t]
\vspace{-.3cm}
\begin{minipage}[t]{0.44\linewidth}
\centering
\vspace{.3cm}
\subfloat[$\mathsf{LTS}(T_{HS})$]{\includegraphics[scale=.56, keepaspectratio, valign=t]{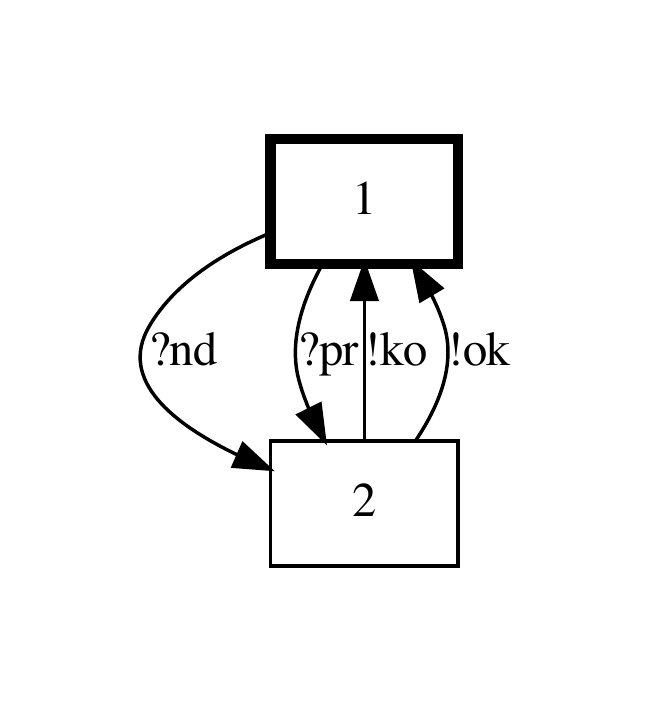}}
\hspace*{0.15cm}
\vspace*{1.75cm}
\caption{Hospital server.}
\label{fig:THSSAut}
\end{minipage}%
\begin{minipage}[t]{0.56\linewidth}
\centering
\subfloat[$\mathsf{LTS}(T''_{HC})$]{\includegraphics[scale=.56,  keepaspectratio, valign=t]{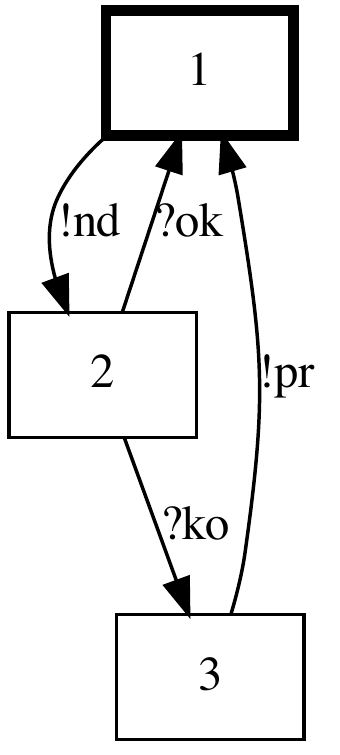}}
\subfloat[$\mathsf{LTS}(T'_{HC})$]{\includegraphics[scale=.56, keepaspectratio, valign=t]{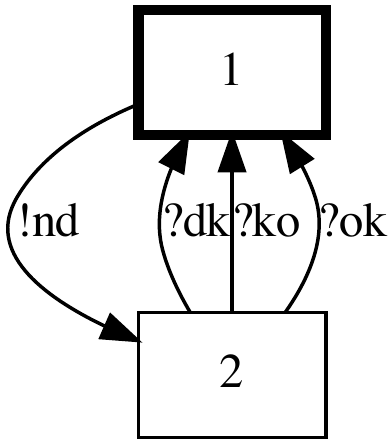}}
\hspace*{0.01cm}
\subfloat[$\mathsf{LTS}(T_{HC})$]{\includegraphics[scale=.56, keepaspectratio, valign=t]{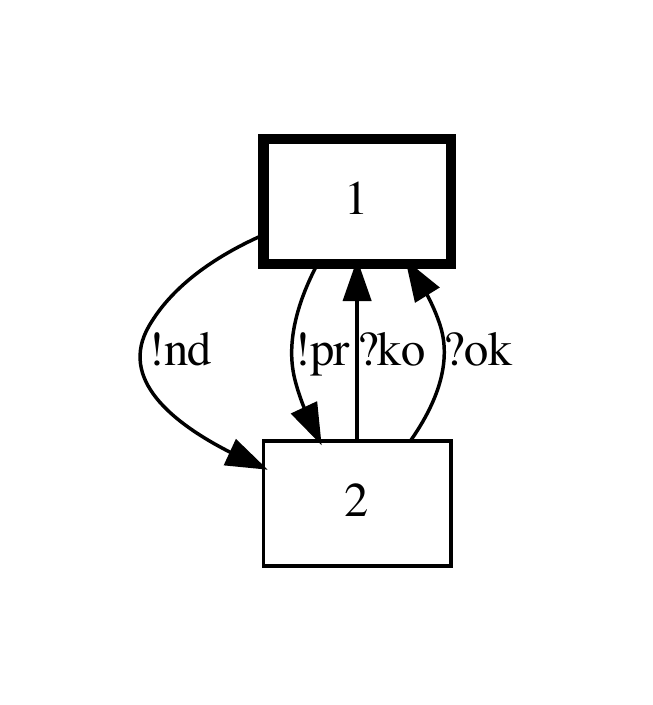}}
\caption{Hospital clients.} 
\label{fig:AsyncAut}
\end{minipage}
\end{figure}

\begin{figure}
\begin{minipage}[t]{0.49\linewidth}
\centering
\vspace*{.3cm}
\subfloat[$\mathsf{LTS}(T_{SS})$]{\includegraphics[scale=.56, keepaspectratio, valign=t]{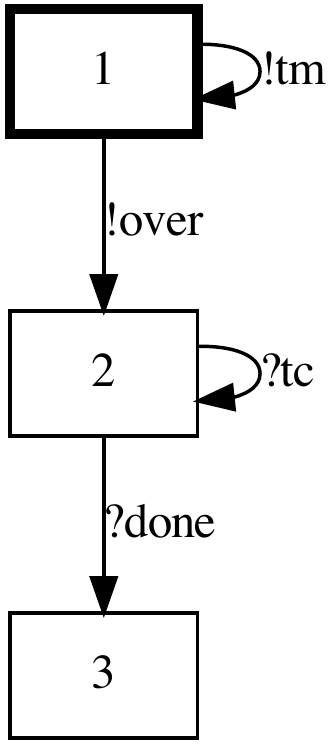}}
\vspace*{.05cm}
\caption{Satellite protocol server.}
\label{fig:SFAsyncAut}
\end{minipage}%
\begin{minipage}[t]{0.49\linewidth}
\centering
\subfloat[$\mathsf{LTS}(T'_{SC})$]{\includegraphics[scale=.56, keepaspectratio, valign=t]{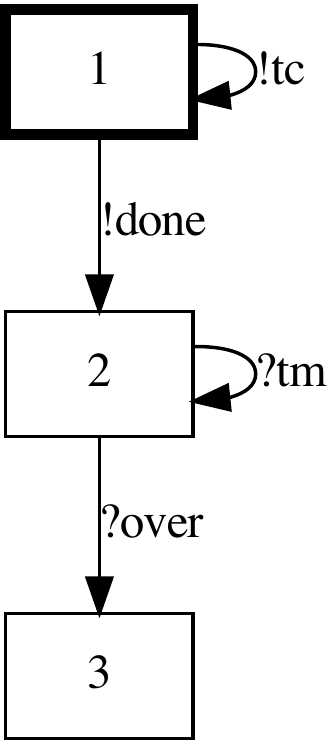}}
\hspace*{.5cm}
\subfloat[$\mathsf{LTS}(T_{SC})$]{\includegraphics[scale=.56, keepaspectratio, valign=t]{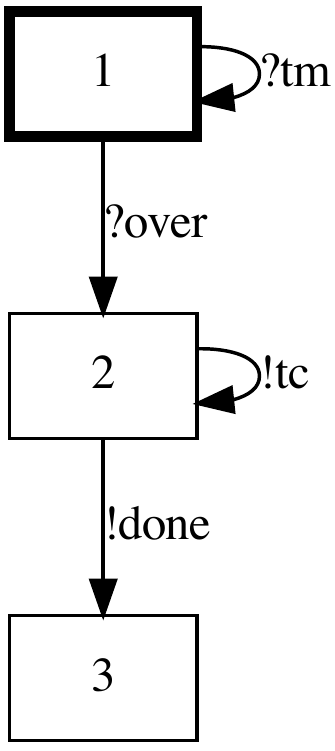}}
\caption{Satellite protocol clients.}
\label{fig:FAsyncAut}
\end{minipage}
\end{figure}

We first recall the syntax of session types
and their automata representation in the style of 
communicating finite-state machines (CFSM) \cite{cfsm83}. We then show how our tool can generate simulation
graphs for supported session subtyping relations:
synchronous \cite{LangeY16}, asynchronous \cite{BCLYZ21} and fair asynchronous subtyping~\cite{BLZ21}.
In the asynchronous cases automata are assumed to communicate over unbounded \textsc{fifo} channels as for CFSMs.

%
%

\subsection{Session Types and their Automata Representation}\label{examples}

%
The formal syntax of two-party session types is given below. Notice that we follow
the simplified notation used in, e.g., \cite{DenielouY13,BravettiCZ17,BLZ21}, 
%
which abstracts away 
from data carried by messages (payloads). This is done in order to focus on the key aspects of the session subtyping problem (as we will see co/contra-variance of output/input and output anticipation): passing data or channels (delegation) are features that we deem orthogonal to such a problem.

\begin{definition}[Session Types]\label{def:sessiontypes}
  Given a set of label names $\mathcal{L}$, ranged over by $l$, the syntax
  of two-party session types is given by the following grammar: \\[.05cm]
\centerline{$    \begin{array}{lrl}
      T\ \ &::=&\ \   
                 \Tselect{l}{T} 
                 \quad \mid\quad  \Tbranch{l}{T} 
                 \quad \mid\quad  \Trec X.T
                 \quad \mid\quad \Tvar X
                 \quad \mid\quad \Tend
    \end{array}$} \\[.05cm]
where $I \!\neq\! \emptyset$ and $\forall i \!\neq\! j \!\in\! I \dst l_i\! \neq\! l_j$.
\end{definition}

%
Type $\Tselect{l}{T}$ represents an internal 
choice among {\it outputs}, specifying that the chosen label name $l_i\! \in\! \call$ is sent and, then,
continuation $T_i$  is executed.
%
$\Tbranch{l}{T}$ represents, instead, an external 
choice among {\it inputs},
specifying that, once a 
label name $l_i\! \in\! \call$ is
received, continuation $T_i$ takes place.
%
%
%
%
%
%
Types $\Trec X.T$ and $\Tvar X$ denote standard recursion
constructs.
%
We assume recursion to be guarded,
i.e., in $\Trec X.T$ the recursion variable $\Tvar X$ 
occurs only after receving or sending a label.
%
%
Type $\Tend$ denotes the end of the interaction.
Session types are closed, 
i.e., 
all recursion variables $\Tvar X$ occur under the scope of a
corresponding binder $\Trec X.T$.  


In the tool we graphically represent the behaviour of a session type $T$ as a Labeled Transition System (LTS), see, e.g., Figures \ref{fig:THSSAut}, \ref{fig:AsyncAut},
\ref{fig:SFAsyncAut} and \ref{fig:FAsyncAut}.
Following the notation of CFSMs, we denote a LTS by $(\States, \state_0, \Trans)$, with $\States$ being a 
set of states, $\state_0$ the initial state and $\Trans$ a transition relation over $\States \!\times\! (\{!,\!?\} \!\times\! \call) \!\times\! \States$,
with label ``$! \, l$'' representing output on $l$ and label ``$? \, l$'' representing input on $l$.

We use $\mathsf{LTS}(T)$ to denote the LTS of type $T$. Let $\calt$ be the set of all session types $T$. 
We define transition relation
$\longrightarrow \; \subseteq \calt \!\times\! (\{!,\!?\} \!\times\! \call) \!\times\!
\calt$, as the least transition set satisfying the following rules\\ 
$
\!\Tselect{l}{T} \!\arrow{ \! ! \, l_i}\! T_i \hspace{.4cm} i \!\in\! I \hspace{.8cm}
 \Tbranch{l}{T} \!\arrow{ \! ? \, l_i}\! T_i \hspace{.4cm} i \!\in\! I \hspace{.8cm}
\infr{ T\{\Trec X.{T}/\Tvar{X}\} \!\arrow{\act}\! T'}{\Trec X.{T} \!\arrow{\act}\! T' } 
$\\
with label $\act$ ranging over $\{!,\!?\} \!\times\! \call$.
Notice that a state $\Tend$, called {\it termination state}, has no outgoing transitions. 
Given a session type $T$ we define $\mathsf{LTS}(T)$ as being $(Q_T,T, 
\Trans_T)$, where:
$Q_T$ is the set of terms $T'$ which are reachable from $T$ according to $\longrightarrow$ relation and $\Trans_T$ is defined as the restriction of $\longrightarrow$ to  $Q_T \!\times\! (\{!,\!?\} \!\times\! \call) \!\times\! Q_T$. 

%

Notice that in general an LTS may express more
behaviours than the ones described by session types: 
it can include non-deterministic and mixed choices, i.e.\ choices including both inputs and outputs. Here we only consider LTSs $(\States, \state_0, \Trans)$ such that $\exists T \!\in\! \calt \ldotp \mathsf{LTS}(T)=(\States, \state_0, \Trans)$.


\begin{example}
As an example of session types we consider the Hospital server from~\cite{BCLYZ21}:\\
\centerline{$
    T_{HS} = \Trec{X}.
                     \outchoicetop \!
                     \left\{
                     \exHQ ; \inchoicetop \{ko; \Tvar{X}, \ ok; \Tvar{X} \}, \
                     \exLQ ; \inchoicetop \{ko; \Tvar{X}, \ ok; \Tvar{X} \}
                     \right\}
$}\\[.05cm]
Figure \ref{fig:THSSAut} shows $\mathsf{LTS}(T_{HS})$ as produced by our tool.
The server $T_{HS}$ expects to receive two types of
messages: $\exHQ$ (next patient data) or $\exLQ$ (patient report).
%
Then it may send either
$\exOK$ or $\exKO$, indicating whether the evaluation of received data
was successful or not, and it loops.
\end{example}

We now define the \textit{dual} of a session type 
$T$, written $\dual{T}$. $\dual{T}$ is inductively
obtained from $T$ as follows:
$\dual{\Tselect{l}{T}} = \Tbranch{l}{\dual{T}}$,
$\dual{\Tbranch{l}{T}} = \Tselect{l}{\dual{T}}$,
$\dual{\Tend} = \Tend$, $\dual{\Tvar{X}} = \Tvar{X}$, and
$\dual{\Trec{X}.T} = \Trec{X}.\dual{T}$. 
For example, 
the dual of the Hospital server $T_{HS}$ is:\\[.05cm]
\centerline{$\dual{T_{HS}} =  \Trec{X}.
                     \inchoicetop \!
                     \left\{
                     \exHQ ; \outchoicetop \{ko; \Tvar{X}, \ ok; \Tvar{X} \}, \
                     \exLQ ; \outchoicetop \{ko; \Tvar{X}, \ ok; \Tvar{X} \}
                     \right\}
$}

\begin{example}\label{ExAsync}
We now consider examples of session types that are clients of the Hospital service: an ``ideal'' client $T_{HC}$ and two specific ones $T'_{HC}$ and  $T''_{HC}$, respectively. \\
\centerline{$
  \begin{array}{lcl}
    T_{HC} & = & \dual{T_{HS}} = \Trec{X}.
                     \inchoicetop \!
                     \left\{
                     \exHQ ; \outchoicetop \{ko; \Tvar{X}, \ ok; \Tvar{X} \}, \
                     \exLQ ; \outchoicetop \{ko; \Tvar{X}, \ ok; \Tvar{X} \}
                     \right\}
                         \\
    T'_{HC} & = & \Trec{X}.
                     \inchoicetop \!
                     \left\{
                     \exHQ ; \outchoicetop \{ko; \Tvar{X}, \ ok; \Tvar{X}, \ dk; \Tvar{X} \}
                     \right\}
                         \\
    T''_{HC} & = & \Trec{X}.  \inchoicetop \! \left\{\exHQ; \outchoicetop \{ ko;  \Tvar{X}, \ ok; \inchoicetop \! \{\exLQ; \Tvar{X} \}\}\right\}
    \\
  \end{array}
$}\\[.05cm]
Figure \ref{fig:AsyncAut} shows $\mathsf{LTS}(T''_{HC})$, $\mathsf{LTS}(T'_{HC})$ and $\mathsf{LTS}(T_{HC})$,\footnote{As we will see, the order in which the LTSs are presented reflects the subtyping relation (we will show that $T''_{HC}$ and $T'_{HC}$ are subtypes of $T_{HC}$) and the positions in which types are inputed in the tool.} as produced by our tool.

The ``ideal'' client  $T_{HC}$ is simply the dual of the Hospital server: first it may send two types of
messages $\exHQ$ or $\exLQ$, then it expects to receive either $\exOK$ or $\exKO$. In general, a client that is {\it compliant} with the Hospital server is a type such that: $(i)$ each message sent by the client (resp.\ server) can be received by the server (resp.\ client), and $(ii)$ neither the server nor the client blocks in a receive.
%
For example, the client $T'_{HC}$, a slightly modified version of $T_{HC}$ that may send $\exHQ$ only 
and expects to receive also $dk$ (don't know) besides $\exOK$ and $\exKO$ (i.e., it applies covariance of outputs and contravariance of inputs, see \cite{LangeY16}) is still compliant with the Hospital server. 
Hence we say that $T'_{HC}$ is a subtype of $T_{HC}$.
%

Under asynchronous communication client compliance is relaxed by requiring that all messages that are sent are {\it eventually} received. For example, in this setting,
client $T''_{HC}$ (that may anticipate output $\exHQ$ w.r.t. inputs) is also a compliant client, see~\cite{BCLYZ21}, hence $T''_{HC}$ is an asynchronous subtype of $T_{HC}$.

Notice that, both for synchronous and asynchronous communication (see~\cite{BCZ18}),  it holds, for any session type $T,T'$: $T'$ subtype of $T$ implies $\dual{T}$ subtype of $\dual{T'}$ (closure under duality). As we will see, our tool automatically handles the generation of the {\it dual subtyping problem} ($\dual{T}$ subtype of $\dual{T'}$) from $T'$ subtype of $T$ by exchanging and dualizing inputted types.

\end{example}
\begin{example}\label{SatEx}
As another example, 
we consider clients of the Satellite protocol from~\cite{BLZ21}: an ``ideal'' client (the dual of the Satellite protocol server $T_{SS}$ whose LTS is depicted in Figure \ref{fig:SFAsyncAut}) and a specific one; here denoted with $T_{SC}$ and $T'_{SC}$, respectively.\\
\centerline{$
  \begin{array}{lcl}
    T_{SC} & = & \dual{T_{SS}} = 
                 \Trec{X}.~\Tbra{ \exTM ; \Tvar{X} ,  \,\exOVER ; \Trec{Y}
                 . \Tsel{ \exTC ; \Tvar{Y}  , \,\exDONE ; \Tend } } 
    \\
    T'_{SC} & = & \Trec  X .\Tsel{ \exTC ; \Tvar X  , \, \exDONE ;
                  \Trec{Y} . 
                  ~\Tbra{ \exTM ; \Tvar{Y} ,  \,\exOVER ; \Tend } } 
  \end{array}
$}\\[.05cm]
Figure \ref{fig:FAsyncAut} shows $\mathsf{LTS}(T'_{SC})$ and $\mathsf{LTS}(T_{SC})$, as produced by our tool. 
The ``ideal'' client  $T_{SC}$ may receive 
a 
number of telemetries ($\exTM$), followed by a
message $\exOVER$. 
%
In the second phase, the client 
sends a number of telecommands
($\exTC$), followed by a message $\exDONE$. Under {\it fair} asynchronous communication  client $T'_{SC}$ (with phases exchanged) is also compliant with the server, i.e.\ $T'_{SC}$ a fair asynchronous subtype of $T_{SC}$, see \cite{BLZ21}. Compared to asynchronous communication considered in Example \ref{ExAsync}, here client compliance entails that,
under {\it fairness assumption} (i.e.\ communication loops with some exit are assumed to be eventually escaped), both the client and the server must reach successfull termination with no messages left to be consumed in the \textsc{fifo} 
channels.
\end{example}

\begin{figure}
\begin{minipage}[t]{0.49\linewidth}
\centering
\subfloat{\includegraphics[scale=.3,  keepaspectratio, valign=t]{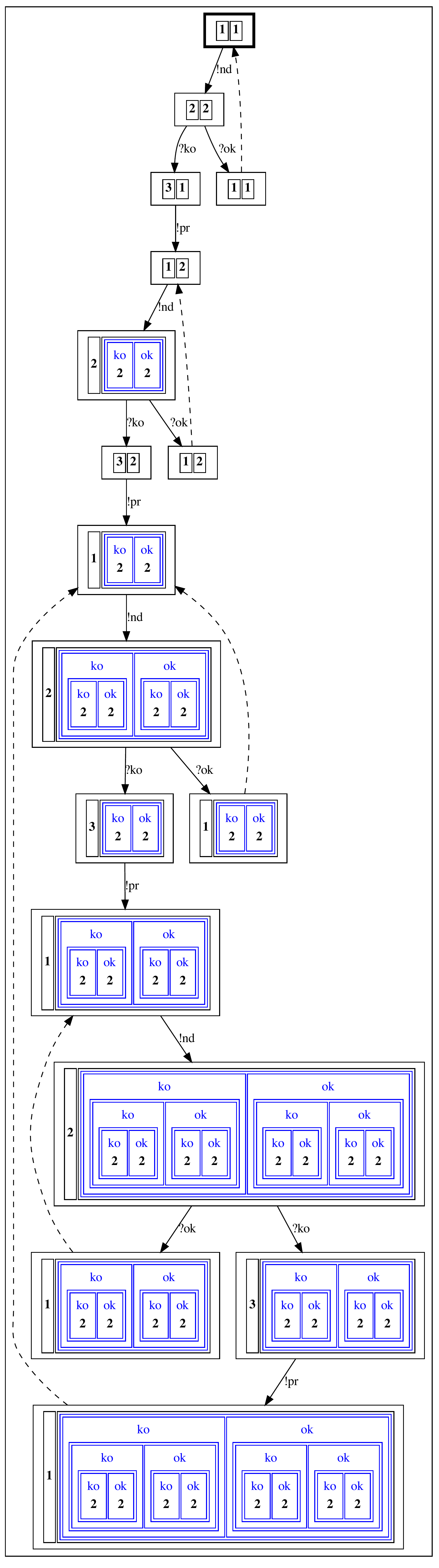}}
\caption{Asynchronous simulation.}
\label{fig:AsyncSim}
\end{minipage}
\begin{minipage}[t]{0.49\linewidth}
\centering
\subfloat{\includegraphics[scale=.7,  keepaspectratio]{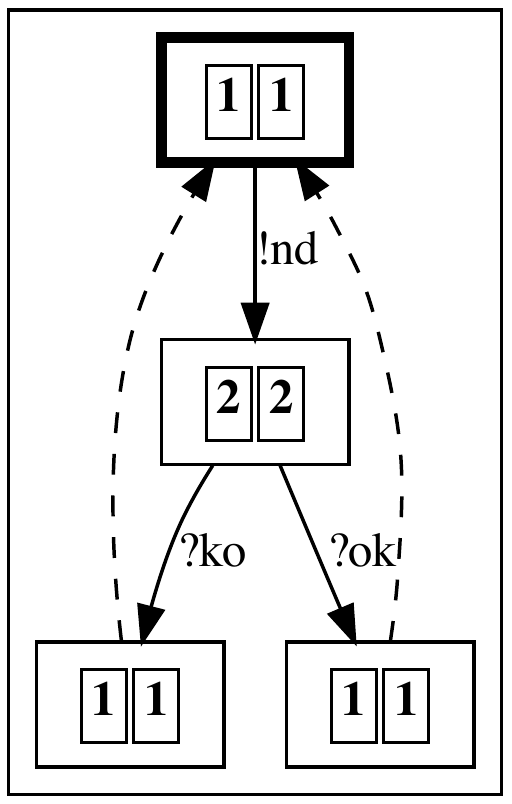}}
\caption{Synchronous simulation.} 
\label{fig:SyncSim}
\subfloat{\includegraphics[scale=.7,  keepaspectratio]{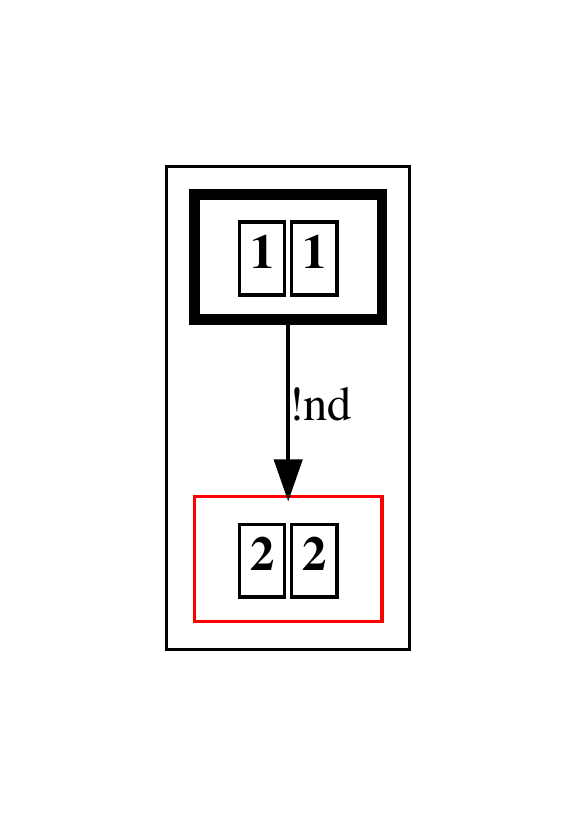}}
\caption{Failed synchronous simulation.} 
\label{fig:FailSyncSim}
\end{minipage}
\end{figure}

\subsection{Synchronous Session Subtyping}\label{sub:syncsub}
In order to establish whether type $T'$ is a synchronous subtype of a type $T$ \cite{LangeY16} we can perform synchronous simulation of the (ordered) pair of LTSs $\mathsf{LTS}(T')=(Q',q'_0,\Trans')$ and $\mathsf{LTS}(T)=(Q,q_0,\Trans)$. Simulation states are pairs $(q',q)$, with $q' \!\in\! Q'$ and $q \!\in\! Q$. The simulation proceeds by starting from state $(q'_0,q_0)$ and by synchronously matching transitions of $\mathsf{LTS}(T')$ and $\mathsf{LTS}(T)$ having the {\it same} labels (both
``$! \, l$'' or both ``$? \, l$''). 
For each reached simulation state $(q',q)$ 
we must have: $(i)$ the set of outputs (resp.\ inputs) fireable by $q'$ is subset (resp.\ superset) or equal to the set of outputs (resp.\ inputs) fireable by $q$; this enacts covariance (resp.\ contravariance) of outputs (resp.\ inputs), $(ii)$ if $(q',q)$ performs no transitions then both $q'$ and $q$ must perform no transitions (successfully terminate). 

On the contrary, simulation states $(q',q)$ for which the above constraints are not satisfied are called \textit{failure simulation states} (depicted in red in our tool) and cause synchronous subtyping not to hold. 
\begin{example}
Figure \ref{fig:SyncSim} shows the synchronous simulation graph, as produced from our tool, for the pair $\mathsf{LTS}(T'_{HC})$ and $\mathsf{LTS}(T_{HC})$. Notice that our tool builds the simulation graph as a tree: when a pair $(q',q)$ is reached, which was previously traversed (as e.g.\ for the $(1,1)$ pair), simulation does not proceed further in that branch and a dashed line is depicted connecting the two copies of $(q',q)$. Notice that, if in $T'_{HC}$ we turn $\rcv{ko}$ into $\rcv{ko1}$ (creating a mismatch with the server), $T'_{HC}$ is no longer a synchronous subtype of $T_{HC}$. This can be seen in Figure \ref{fig:FailSyncSim} where the originated failure simulation state is depicted in red.
\end{example}
We now give the formal definition of synchronous subtyping. We first define set of inputs and set of outputs fireable by a state $q$ as follows:  $\inTrans{q}=\{l \mid \exists q'.q\trans{?\, l} q'\}$ and
  $\outTrans{q}=\{l \mid \exists q'.q\trans{!\, l} q'\}$.

\begin{example}
  Consider $\mathsf{LTS}(T_{HC})$ (Figure~\ref{fig:AsyncAut}), we have the following:
  \[
    \begin{array}{lllll}
      \inTrans{1} = \emptyset
      &
        \hspace{.3cm}\inTrans{2}=  \{ \exKO, \exOK \}
      \\
       \outTrans{1} = \{ \exHQ, \exHQ\}
      &
        \hspace{.3cm}\outTrans{2} = \emptyset
    \end{array}
  \]
\end{example}

\begin{definition}[Synchronous Simulation]\label{def:syncsimtree}
Given set of label names $\call$ and two
LTSs $(P,p_0,\Trans_1)$ and $(Q,q_0,\Trans_2)$,
synchronous simulation is defined as a labeled transition system over states of $P \!\times\! Q$, i.e. pairs denoted by $\simtreepair pq$, with $p \!\in\! P$ and $q \!\in\! Q$. In particular, the initial state is $\simtreepair{p_0}{q_0}$ and the transition relation $\simtreetrans{}$, labeled over  $\{!,\!?\} \!\times\! \call$,
is defined as the minimal relation satisfying rules:\\[.1cm]
$    \begin{array}{c}
      \infer[\!\mathsf{(In)}]
      {\simtreepair pq\ \simtreetrans{? \,l}\ \simtreepair{p'}{q'}}
      {p\trans{? \,l}_1 p' \;\; q\trans{? \,l}_2 q' \;\; \inTrans{p} \!\supseteq\! \inTrans{q}}      
      \quad\quad
      \infer[\!\mathsf{(Out)}]
                                         { \simtreepair pq\ \simtreetrans{!\,l}\ \simtreepair{p'}{q'} }
      {p\trans{! \,l}_1 p' \;\; q\trans{! \,l}_2 q' \;\; \outTrans{p} \!\subseteq\! \outTrans{q}}
      \\
    \end{array}
$
\end{definition}

Formally, a type $T'$ is a synchronous subtype of a type $T$ if the LTS obtained as the synchronous simulation of the pair $\mathsf{LTS}(T')$ and $\mathsf{LTS}(T)$ is such that, for every state $\simtreepair{q'}{q}$ reachable from the initial simulation state, we have: if $\simtreepair{q'}{q}$ performs no transitions then both $q'$ and $q$ perform no transitions.

\subsection{Asynchronous Session Subtyping}\label{sub:asyncsub}

In contrast to synchronous simulation, the asynchronous one gives the possibility of ``anticipating'', in the right-hand LTS, output transitions w.r.t.\ input transitions that precede them. This can be shown, using our tool, via an example. 
\begin{example}\label{exAsync}
Figure~\ref{fig:AsyncSim} shows the asynchronous simulation tree, as produced from our tool, for the pair $\mathsf{LTS}(T''_{HC})$ and $\mathsf{LTS}(T_{HC})$. Simulation of Figure \ref{fig:AsyncSim} proceeds as follows.
For instance, after transitions $\snd{\exHQ}$, $\rcv{ko}$ and $\snd{\exLQ}$ (i.e.\ path ``$\snd{\exHQ} \, \rcv{ko} \, \snd{\exLQ}$'') are synchronously performed by $\mathsf{LTS}(T''_{HC})$ and $\mathsf{LTS}(T_{HC})$, they reach states $1$ and $2$, respectively. Now, $\mathsf{LTS}(T''_{HC})$ in state $1$ can only do output $\snd{\exHQ}$, while $\mathsf{LTS}(T_{HC})$ in state $2$ can only do inputs. Being asynchronous, the simulation can proceed by calculating the so-called state $2$
{\it input tree} $\intree{2} = \echoice{\exKO \!:\! 1 , \, \exOK \!:\! 1}{}$, i.e.\ the spanning tree from state $2$ (constructed considering input transitions only), which has two leaves, both being state $1$. 
Provided that all leaves 
of 
$\intree{2}$ can perform $\snd{\exHQ}$, the simulation can proceed by considering $\echoice{\exKO \!:\! [\,] , \, \exOK \!:\! [\,]}{}$ as an ``accumulated'' input for the right-hand LTS and by making all states in its leaves evolve by performing the $\snd{\exHQ}$ transition. Therefore, after simulation performs 
$\snd{\exHQ}$, $\mathsf{LTS}(T''_{HC})$ and $\mathsf{LTS}(T_{HC})$ reach states $2$ and $\echoice{\exKO \!:\! 2 , \, \exOK \!:\! 2}{}$, respectively (the state reached by the right-hand LTS 
is actually an input tree). 
\end{example}

In general, input trees (e.g. $\echoice{\exKO : 2 , \, \exOK : 2}{}$ in the example above) are defined in \cite{BCLYZ21} 
as {\it input contexts} ${\mathcal A}$ (representing ``accumulated'' input, e.g. $\echoice{\exKO : [] , \, \exOK : []}{}$ in the example above) with {\it holes} ``$[\,]$'' replaced by LTS states. Their syntax is:
\\
\centerline{$ {\mathcal A}\ \ ::=\ \ [\,] \ \mid\ \echoice{l_i:{\mathcal
      A}_i}{i\in I} $}\\
In the tool we represent input trees by nested boxes. For instance the input tree
$\echoice{\exKO : \echoice{\exKO : 1 , \, \exOK : 1}{} , \, \exOK : \echoice{\exKO : 1 , \, \exOK : 1}{} }{}$ is represented as:\\[.2cm] 
\centerline{\includegraphics[scale=.8, keepaspectratio]{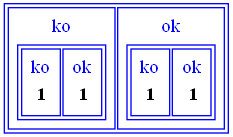}}\\
\vspace*{-.4cm}

In general, due to input accumulation, represented by an input tree,  even if two types are in an asynchronous subtyping relation, the simulation could proceed infinitely without meeting failure simulation states (as it would happen for the pair $\mathsf{LTS}(T''_{HC})$ and $\mathsf{LTS}(T_{HC})$ of Example \ref{exAsync}).
In our tool we use the algorithm of \cite{BCLYZ21} for checking asynchronous subtyping, which is sound but not complete (in some cases it terminates without returning a decisive verdict). In a nutshell, such an algorithm proceeds as follows. The subtyping simulation terminates when we encounter a failure state (depicted in red in our tool), meaning that the two types are not in the subtyping relation, or when we detect a repetitive behaviour in the simulation (which, we show, can always be found, in case of infinite simulation).  In the latter case, we check whether this repetitive behaviour satisfies sufficient conditions (see \cite{BCLYZ21} for details) that guarantee that the subtyping simulation will never encounter failures. If the conditions are satisfied the algorithm concludes that the two types are in the subtyping relation, otherwise a \textit{maybe} verdict is returned. 

Therefore, the tool always produces a finite simulation tree: for types that are detected to be subtypes, simulation can stop in a state that is identified (via a dashed transition in our tool) to a previously encountered state, even if they are not identical; see bottommost state of Figure \ref{fig:AsyncSim} (outgoing dashed transition). The sufficient conditions checked by the algorithm guarantees that the behaviour beyond such a simulation state is a repetition of the behaviour already observed.

We now give the formal definition of asynchronous subtyping. Given a LTS $(\States, \state_0, \Trans)$,
we write $\state_0 \trans{\act_1 \cdots \act_k} \state_k$ iff there
are $\state_1, \ldots , \state_{k-1} \in \States$ such that
$\state_{i-1} \trans{\act_i} \state_i$ for $1 \leq i \leq k$.
Given a list of messages $\word = \msg{l}_1 \cdots \msg{l}_k$ ($k \geq
0$), we write $\rcvL{\word}$ for the list $\rcv{l}_1 \cdots \rcv{l}_k$
and $\sndL{\word}$ for $\snd{l}_1 \cdots \snd{l}_k$.


\begin{definition}[Input Context]
  An input 
context is a term of the grammar\\[.1cm]
\centerline{$ {\mathcal A}\ \ ::=\ \ [\,]_j \ \mid\ \echoice{l_i:{\mathcal
      A}_i}{i\in I} $}\\[.1cm] 
where: All indices $j$, denoted by
  $I(\mathcal A)$, are distinct and are associated to holes. Moreover, $I \!\neq\! \emptyset$ and $\forall i \!\neq\! j \!\in\! I \dst l_i\! \neq\! l_j$.
\end{definition}
Holes are, thus, actually indexed so to make it possible to individually replace them. In this way ${\mathcal A}[q_i]^{i \in I(\cala)}$ denotes the {\it input tree} obtained by syntactically replacing each hole $[\,]_i$ in
$\ctx A$ by a specific state $q_i \in Q$. 
%
In the sequel, we use $\cali\calt_Q$ to denote the set of input trees over states
$q \in Q$.



\subsubsection{Auxiliary functions}
%
%
Given a CSFM $(\States, \state_0, \Trans)$ and a state
$q \in \States$, we define:
\vspace{-.3cm}
\begin{itemize}
\item
  $\loopio{\star}{q} \iff \exists \word \in \call^{*}, \word' \in
  \call^{+}, q' \in Q \dst q \trans{\star {\word}} q' \trans{\star
    {\word'}} q'$ (with $\star \in \{!,?\}$),
\item the \emph{partial} function $\intree{\cdot}$ as\\
$\intree{q} =
\begin{cases}
  \perp & \text{if }  \loopio{?}{q}
  \\
  q & \text{if } \inTrans{q} = \varnothing
  \\
\echoice{l_i: \intree{q'_i}}{i\in I} & \text{if }   \inTrans{q} = \{l_i \qst i\! \in\! I\} \neq  \varnothing
\end{cases}$\\
with $q'_i$ being the state such that $q\trans{?\, l_i} q'_i$.
\end{itemize}
Predicate $\loopio{\star}{q}$ says that, from $q$, we
can 
reach a cycle with only sends (resp.\ receives), depending on whether
$\star=!$ or $\star=?$.
%
%
%
%
The partial function $\intree{q}$, when defined,  returns the tree containing all sequences
of messages which can be received from $q$ until a final or sending
state is reached.
Intuitively, $\intree{q}$ is undefined when $\loopio{?}{q}$ as it
would return an infinite tree.

\begin{example}
  Consider $\mathsf{LTS}(T_{HC})$ (Figure~\ref{fig:AsyncAut}), we have the following:
 \vspace*{-0.5cm}\begin{figure}[h]
 \begin{minipage}[t]{0.65\linewidth}
  \[
    \begin{array}{lllll} 
      \\
      \intree{1} = 1
      &   \hspace{.3cm}\intree{2} = \echoice{\exKO : 1 , \,
                          \exOK : 1}{}

    \end{array}
  \]
\end{minipage}
\begin{minipage}[t]{0.3\linewidth}
\centering
\includegraphics[width=0.4\linewidth, valign=t]{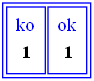}
\\[.1cm]{\!\!$\intree{2}$ tool representation}
\label{intree}
\end{minipage}%
\end{figure}
\end{example}

\begin{figure}[t]
\begin{minipage}[t]{0.4\linewidth}
\centering
\subfloat{\includegraphics[scale=.62, keepaspectratio]{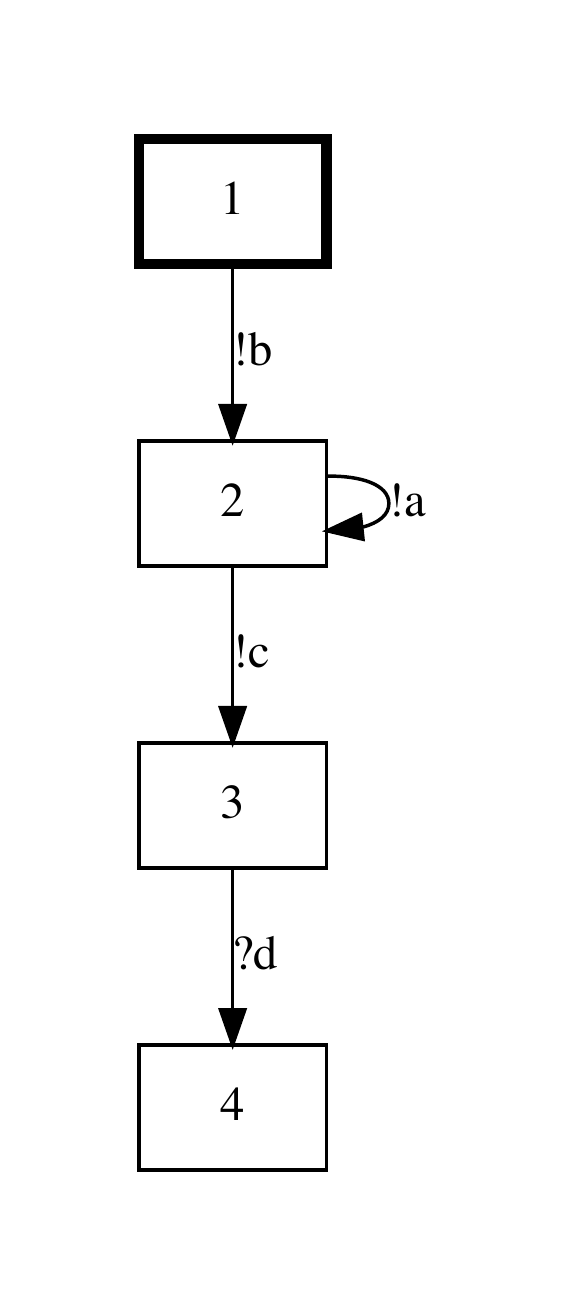}}
\caption{Output cycle example.} 
\label{fig:CycleAut}
\end{minipage}
\begin{minipage}[t]{0.6\linewidth}
\centering
\subfloat{\includegraphics[scale=.4, keepaspectratio, valign=t]{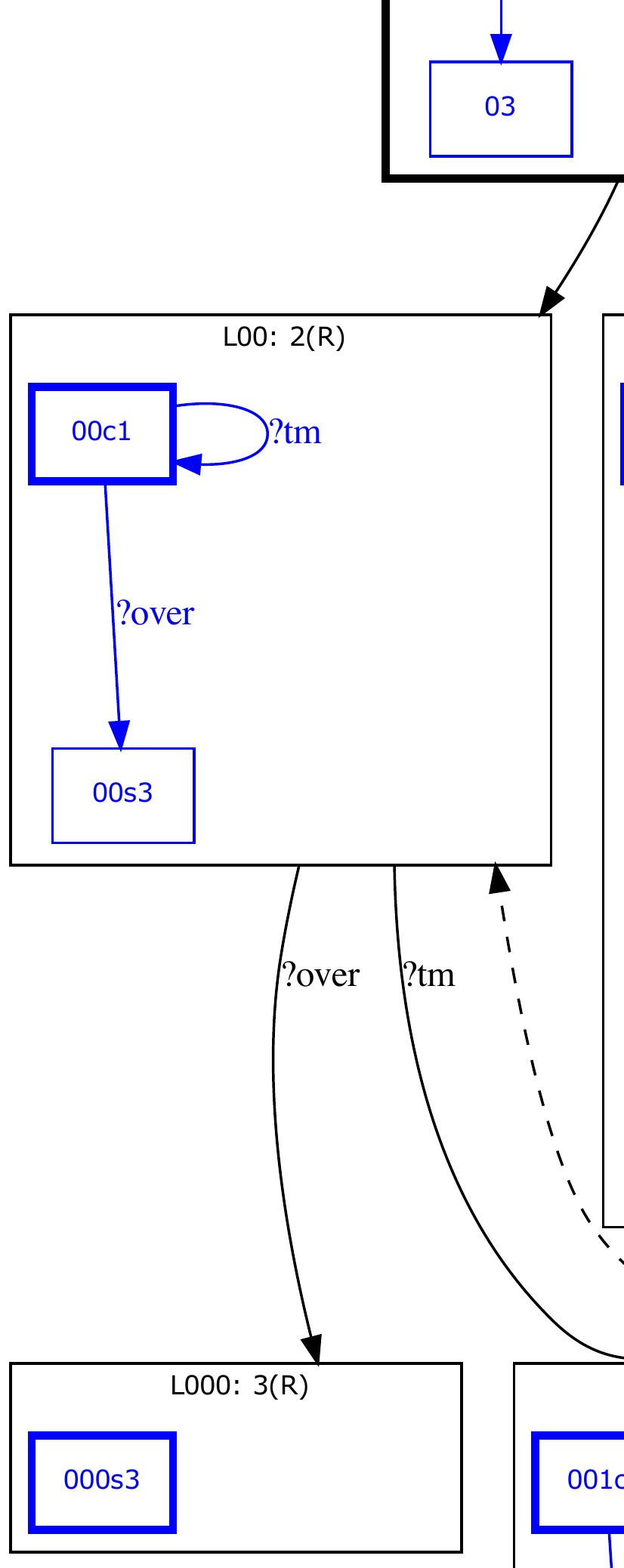}}
\caption{Fair asynchronous simulation fragment.}
\label{fig:FAsyncSim}
\end{minipage}
\end{figure}

\begin{example}\label{example:newexample}
  Consider the LTS of Figure \ref{fig:CycleAut}.
  From state $1$ we can reach state $2$ with an output. The latter can loop with an output into itself. Hence, we have both $\loopio{!}{1}$ and
  $\loopio{!}{2}$.
\end{example}

\begin{definition}[Asynchronous Simulation]\label{def:simtree}
Given set of label names $\call$ and two
LTSs: LTS$_1\!=\!(P,p_0,\Trans_1)$ and LTS$_2\!=\!(Q,q_0,\Trans_2)$,
asynchronous simulation is defined as a labeled transition system over states of $P \!\times\! \cali\calt_Q$, i.e. pairs denoted by $\simtreepair{p}{\ctx A[q_j]^{j\in J}}$, with $p \!\in\! P$ and $\ctx A[q_j]^{j\in J} \!\in\! \cali\calt_Q$. In particular, the initial state is $\simtreepair{p_0}{q_0}$ and the transition relation $\simtreetrans{}$, labeled over  $\{!,\!?\} \!\times\! \call$,
is defined as the minimal relation satisfying rules in Definition \ref{def:syncsimtree}, plus the following ones:\\[.1cm]
$    \begin{array}{c}
	\infer[\!\mathsf{(InCtx)}]
      {
      \simtreepair{p}{ \echoice{l_i:\ctx A_i[q_{i,j}]^{j\in J_i}}}{i\in I} 
      \ \simtreetrans{?\, l_k}\ 
      \simtreepair{p'}{\ctx A_k[q_{k,j}]^{j\in J_k}}} 
      {p \trans{?\, l_k}_1 p' &  k\in I &
                                      \inTrans{p} \supseteq \{l_i \mid
                                      i\in I\, \}}
      \\[4mm]
      \infer[\!\!\mathsf{(OutA)}]
      {
      \simtreepair p{\ctx A[q_j]^{j\in J}}
      \simtreetrans{!\, l}\simtreepair{p'}
      {\ctx A[\ctx A_j[q'_{j,h}]^{h\in H_j}]^{j\in J}}
      }
      {
      \begin{array}{c@{}}
        \!
        p\trans{!\, l}_1 p'
        \qquad\qquad
        \neg\loopio{!}{p}
        \\
        \!\forall j \!\!\in\!\! J. \big(
        \intree{q_{j\!}}\! =\! \ctx A_{j\!}[q_{j,h}]^{h\in H_j} \!\!\land\!
        \forall h \!\in\!\! H_j.
        ( \outTrans{p} \!\subseteq\! \outTrans{q_{j,h}} \!\land\!
        q_{j,h\!} \!\trans{!\, l}_2 \! q'_{j,h}) \big)
      \end{array}
      }
    \end{array}
$

\end{definition}
The two additional rules express how inputs are accumulated and consumed by means of input trees in the LTS$_2$.
The first one is applicable when the input tree state of the LTS$_2$ is
non-empty and the state $p$ of the LTS$_1$ is able to perform a
receive action corresponding to any message located at the root of the
input tree (contra-variance of receive actions).
The second rule allows the LTS$_1$ to execute some send
actions by matching them with send actions that, in the LTS$_2$, occur after receives.
Intuitively, each send action outgoing from state $p$ of the LTS$_1$ must also be
executable from each of the states $q_{j,h}$ of $\intree{q_{j\!}}\! =\! \ctx A_{j\!}[q_{j,h}]^{h\in H_j}$, with
$q_{j\!}$ being a leaf of the input tree state ${\ctx A[q_j]^{j\in J}}$ of the LTS$_2$ (covariance of send actions).
The constraint $\neg\loopio{!}{p}$ guarantees that accumulated receive actions will be eventually executed.




\subsection{Fair Asynchronous Session Subtyping}\label{sub:fairasyncsimtree}

Consider again the satellite protocol of Example \ref{SatEx}. The asynchronous subtyping of previous Section \ref{sub:asyncsub} rejects $T'_{SC}$ as a
subtype of $T_{SC}$.
Indeed that notion of subtyping allows for anticipation of outputs
only when they are preceded by a \emph{bounded} number of inputs.
However the outputs of $T_{SC}$ occur after an arbitrary
number of inputs.
That notion of subtyping requires that all sent messages are consumed
along \emph{all} possible computations of the receiver.
While in $T'_{SC}$ there is a degenerate execution where
the candidate subtype sends an infinite number of $\mathsf{tc}$
messages and thus never performs the required inputs.

In contrast, {\it fair} asynchronous session subtyping~\cite{BLZ21} relies on the
assumption that such degenerate executions cannot occur under the
natural assumption the loop of outputs eventually terminates, i.e.,
only a finite (but unspecified) amount of messages can be emitted.

Concretely, the fair subtyping uses a more expressive notion of input
contexts ${\mathcal A}$ that also include recursive constructs.
Their syntax becomes:
\\
\centerline{$ {\mathcal A}\ \ ::=\ \ [\,] \ \mid\ \echoice{l_i:{\mathcal
      A}_i}{i\in I} \ \mid\ \Trec X.{\mathcal A}  \ \mid \ 
    \Tvar X 
    $}\\
  These input context can encode the recursive reception of
  messages in the satellite example and thus identify
  $T'_{SC}$ as a fair asynchronous subtype of
  $T_{SC}$.

Figure~\ref{fig:FAsyncSim} shows a fragment of the resulting fair
asynchronous simulation tree, as produced from our tool: due to the
more complex syntax of input contexts, states now contain a (possibly
looping) automaton instead of an input tree.

%
%


\section{Main Functionalities of the Tool} \label{SEC:tool}


Besides the classic operations that a text editor allows (e.g.\ edit, load, save), users can compute the dual of: either a single session type or the entire subtyping problem. To facilitate understanding of session types, the tool offers the possibility to view/save the graphical representation of a given type by means of “Show Image” and “Save Image” respectively.  In our tool types are inputted by means of {\it two text areas}: the leftmost one is used for the candidate subtype and the rightmost one for the candidate supertype. Input types must be expressed with the syntax presented in Definition \ref{def:sessiontypes},  with ``$+$'' standing for ``$\oplus$'' and ``$rec$'' standing for ``$\mu$''. In addition the tool accepts: $(i)$ the alternative ``raw'' syntax $[! \, a \, ; T , ! \, b \, ; T' , \dots]$ standing for $\oplus \{a \, ; T , b \, ; T' , \dots \}$
and $[? \, a \, ; T , ? \, b \, ; T' , \dots]$ 
for $\& \{a \, ; T , b \, ; T' , \dots \}$ $(ii)$ the abbreviations $! \, a \, ; T$ and $? \, a \, ; T$ standing for $\oplus \{a \, ; T \}$ and $\& \{a \, ; T \}$.  A Python parser checks that the inputted types fit the above syntax using the following \textit{EBNF} syntax: \\[.1cm]
$
  \begin{array}{lcl}
    S &::=&  \,OP\, \textbf{\{}\,\textit{\textbf{id}}\, \textbf{;} \,S\, (\textbf{, } \,\textit{\textbf{id}}\, \textbf{;} \,S\,)^*\textbf{\} } \ \mid\ \ \textbf{rec} \,\textit{\textbf{id}}\,\textbf{. }\,S\,\ \mid\ \ \,\textit{\textbf{id}}\, \ \mid\ \ \textbf{end}
     \\
    && \textbf{!} \,\textit{\textbf{id}}\, \textbf{;} \,S\,  \ \mid\ \  \textbf{[ !} \,\textit{\textbf{id}}\, \textbf{;} \,S\, (\textbf{, !} \,\textit{\textbf{id}}\, \textbf{;} \,S\,)^* \textbf{]}  \ \mid \ \ \textbf{?} \,\textit{\textbf{id}}\, \textbf{;} \,S\, \ \mid\ \ \textbf{[ ?} \,\textit{\textbf{id}}\, \textbf{;} \,S\, (\textbf{, ?} \,\textit{\textbf{id}}\, \textbf{;} \,S\,)^* \textbf{]}
  \\
  OP& ::=& \textbf{+} \ \mid\ \ \textbf{\&}
  \end{array}
$\\[.2cm]
where \textit{\textbf{id}} is a non-empty sequence of uppercase and lowercase letters possibly followed by trailing numbers.

The core of the tool is the algorithm menu.  Users can choose between different subtyping algorithms and possibly set a maximum number of  execution steps. The algorithm response can be: “true", “false" or “maybe" (for asynchronous algorithms, due to undecidability, or when the specified number of steps 
is not enough to determine the subtyping relation), along with the time needed.  In addition,  it is possible to run all the algorithms to have an overview 
of the types of relationship that hold. 
Finally, the “Simulation Result” menu, which is initially disabled, makes it possible to show or save the graphical output of the last performed algorithm.

\subsection{Extensibility of the Tool}\label{extens}

Our tool (and its GUI) is automatically extensible with new subtyping algorithms by simply modifying its json configuration file that we will detail in the next section. Such a file can also be modified, directly from the GUI, by using the ``Algorithm configuration” menu under ``Settings”.  Configuration of a new algorithm is done by providing: its displayed name and the path and calling pattern of its execution command, in the form \\
\centerline{\texttt{ExecutableName [flag] [t1] [t2] [steps]}}\\
The tool will replace [flag][t1][t2][steps] with: the user-selected flags, the pair of session types the user wants to check and the number of steps the algorithm is requested to do. The above order of bracketed elements ([steps] is optional) may change according to the algorithm. The json file also maps algorithm-dependent flag names into tool functionalities by categorizing them. For instance, the default \textit{flag} category includes flags that simply modify the behaviour of algorithms: e.g.\ the asynchronous one has the -\;\!-\;\!$\mathsf{nofallback}$ flag that prevents the algorithm from trying to fall back to the dual subtyping problem in case of an initial maybe verdict. Moreover,
the \textit{execution flag} category is useful when an executable encloses different (alternative) algorithms, e.g.\ Gay and Hole (-\;\!-\;\!$\mathsf{gayhole}$) and Kozen et al.'s (-\;\!-\;\!$\mathsf{kozen}$) algorithms for synchronous subtyping (with one indicated as being the default). Morever, the \textit{visual flag} category includes just the name of the flag causing the algorithm to produce the graphical simulation. 

When adding an algorithm to the tool, the following requirements have to be satisfied: they have to support command line execution (with the possibility of taking .txt files as input) and have to fit the ``raw” syntax described above. Regarding the algorithm response, the only requirement is that it is printed on the standard output. Finally, to generate the graphical output, it is mandatory that the algorithm creates a .dot file no matter what its name is (since it is specified in dedicated section of the json configuration). It is important to observe that our tool is agnostic to the implementation language of algorithms, since it makes use of their executable version. 

\subsection{Configuration of Tool Algorithms}
The json file presented below is an example of the ``algorithms\_config.json" currently used by the tool. The \textit{standard\_exec} field specifies the default 
execution flag, e.g.\ Gay and Hole or Kozen.  Moreover, the \textit{simulation\_file} field indicates the relative path to the algorithm generated \textit{Graphviz} “.dot" simulation file. Similarly \textit{win, osx} and \textit{linux} point at the folder in which the tool looks for the algorithm binaries for that specific os. 
\begin{lstlisting}[language=json,firstnumber=1]
[{
   "alg_name": "Async Subtyping",
   "flag": "--nofallback",
   "execution_flag": "",
   "standard_exec": "",
   "visual_flag": "--pics",
   "simulation_file": "tmp/simulation_tree",
   "win": "asynchronous-subtyping\\win\\",
   "osx": "asynchronous-subtyping/osx/",
   "linux": "asynchronous-subtyping/linux/",
   "exec_comm": "Checker [flags] [t1] [t2]"
 },
 {
   "alg_name": "Fair Async Subtyping",
   "flag": "",
   "execution_flag": "",
   "standard_exec": "",
   "visual_flag": "--debug",
   "simulation_file": "tmp/simulation_tree",
   "win": "fair-asynchronous-subtyping\\win\\",
   "osx": "fair-asynchronous-subtyping/osx/",
   "linux": "fair-asynchronous-subtyping/linux/",
   "exec_comm": "Checker [flags] [t1] [t2] [steps]"
 },
 {
   "alg_name": "Sync Subtyping",
   "flag": "",
   "execution_flag": "--gayhole,--kozen",
   "standard_exec": "--gayhole",
   "visual_flag": "--pics",
   "simulation_file": "tmp/simulation_tree",
   "osx": "sync_subtyping/osx/",
   "win": "sync_subtyping\\win\\",
   "linux": "sync_subtyping/linux/",
   "exec_comm": "Checker [flags] [t1] [t2]"
 }]
\end{lstlisting}

\section{Conclusion}\label{SEC:conc}
In this paper we introduced an integrated extensible GUI-based tool which: applies algorithms for 
synchronous and (fair) asynchronous session subtyping, and 
generates graphical simulations showing how underlying algorithms work. 

Concerning future work, we plan to use our synchronous subtyping simulation algorithm (with error detection) in the context of type checking for object oriented programming languages where classes are endowed with usage protocols~\cite{BravettiFGHJKR20}. Indeed, extending the theory of \cite{BravettiFGHJKR20} with protocol subtyping, would make it possible to verify correctness also for class inheritance.
In particular, we have started integrating our algorithm (see Appendix \ref{integration}) into the Java checker \cite{MGR21}, which is based on~\cite{BravettiFGHJKR20}.
Finally, we plan to extend the syntax of session types managed by our tool, e.g.\ by including passing of data/channels and, possibly, by also encompassing pre-emption mechanisms~\cite{BravettiZ09,Bravetti21}, which are often used in
communication protocols.

\bibliographystyle{abbrv}
\bibliography{biblio,async}

\appendix

\section{Appendix}

\subsection{Synchronous Subtyping Integration}\label{integration}

In the context of a cooperation with Marco Giunti, Joao Mota and professor Antonio Ravara we are currently working on integrating the synchronous subtyping algorithm (that, for this purpose, is being rewritten in the Kotlin programming language) into the Java Type Checker of \cite{MGR21}, so to also verify correctness in the case of class inheritance.
In particular, the idea is to invoke such an algorithm every time we have a class inheritance declaration, with both the subclass and the superclass being endowed with a usage protocol (in the form of Labeled Transition Systems as the ones considered in this paper), describing order of method calls. The only difference between the synchronous algorithm of this paper and the one we aim to integrate in \cite{MGR21} is that in the latter we need to also take into account {\it droppable states}: states having outgoing input transition(s) where the protocol can optionally terminate (or continue by receiving method invocations). More precisely, we aim at producing a first release of the Java Type Checker of \cite{MGR21} that deals with class inheritance with the following limitations: currently up/down-castings are allowed only if the object to be casted is either in the initial or end state (droppable states are excluded). Overcoming such limitations in future releases will require major changes to the basic structure/architecture/functioning of the Java Type Checker in that we have to consider that calling a method of an object will produce consequences not only in its declared static class but also in the run-time one.

\end{document}